# Semi-Supervised Bone Marrow Lesion Detection from Knee MRI Segmentation Using Mask Inpainting Models

Shihua Qin[a,b], Ming Zhang[b], Juan Shan[c], Taehoon Shin[a,d], Jonghye Woo[a], Fangxu Xing*[a],
[a]Gordon Center for Medical Imaging, Harvard Medical School and Massachusetts General Hospital, Boston, MA, USA 02114; [b]Department of Computer Science, Boston University, Boston, MA, USA 02215; [c]Department of Computer Science, Seidenberg School of CSIS, Pace University, New York, NY, USA 10038; [d]Division of Mechanical and Biomedical Engineering, Ewha Womans University, Seoul, South Korea

## ABSTRACT

Bone marrow lesions (BMLs) are critical indicators of knee osteoarthritis (OA). Since they often appear as small, irregular structures with indistinguishable edges in knee magnetic resonance images (MRIs), effective detection of BMLs in MRI is vital for OA diagnosis and treatment. This paper proposes a semi-supervised local anomaly detection method using mask inpainting models for identification of BMLs in high-resolution knee MRI, effectively integrating a 3D femur bone segmentation model, a large mask inpainting model, and a series of post-processing techniques. The method was evaluated using MRIs at various resolutions from a subset of the public Osteoarthritis Initiative database. Dice score, Intersection over Union (IoU), and pixel-level sensitivity, specificity, and accuracy showed an advantage over the multiresolution knowledge distillation method—a state-of-the-art global anomaly detection method. Especially, segmentation performance is enhanced on higher-resolution images, achieving an over two times performance increase on the Dice score and the IoU score at a 448x448 resolution level. We also demonstrate that with increasing size of the BML region, both the Dice and IoU scores improve as the proportion of distinguishable boundary decreases. The identified BML masks can serve as markers for downstream tasks such as segmentation and classification. The proposed method has shown a potential in improving BML detection, laying a foundation for further advances in imaging-based OA research.

## 1. DESCRIPTION OF PURPOSE

Bone marrow lesions (BMLs) are painful alterations in subchondral bone due to repetitive microdamage at the articular surface. They serve as key structural indicators for the onset and progression of knee osteoarthritis (OA)[1,2]. Accurate detection of BMLs is crucial for the treatment and prevention of knee OA. In knee magnetic resonance imaging (MRI), BMLs often appear in irregular shapes and vary widely in size among different subjects, making manual labeling a striving effort compared to labeling other structures. In addition, BMLs within the femur bone visually resemble structured noise in knee MRI, which further increases difficulty in achieving manual segmentations in data of large populations[3].

Anomaly detection, a technique that identifies unexpected and abnormal items that deviate from normally distributed samples, has been widely used in medical imaging. Chatterjee et al.[4] introduced an unsupervised anomaly detection (UAD) pipeline for detecting abnormalities in brain MRI using a context-encoder variational autoencoder (ceVAE)[5], where a VAE was trained on healthy images with a context-encoder restoring masked images. To segment retinal fluid in optical coherence tomography images, the first Generative Adversarial Network (GAN)-based system, AnoGAN[6], was proposed for unsupervised anomaly detection, which introduced a novel inverse-mapping scheme that mapped high-dimensional normal images to a latent space representation. Most generative model-based UAD systems for medical imaging, such as ceVAE and AnoGAN, have limited ability to generate high-resolution medical images due to data limitations and computational expense. Salehi et al.[7] proposed the Multiresolution Knowledge Distillation (MKD) method for anomaly detection. Instead of generating healthy images, MKD enables better transfer of knowledge from the pre-trained expert network to the cloner network, focusing on features that distinguish normality from anomalies. Another method for generating healthy images is mask inpainting, which is a powerful image restoration technique used to remove unwanted structural elements from images. This principle can be effectively applied to medical imaging

---

*E-mail: fxing1@mgh.harvard.edu. Tel: +1 (617) 726-6904.

where anomalies are distributed in a noisy pattern. Instead of generating a complete image, mask inpainting focuses on reconstructing specific regions of interest (ROI) based on the context outside the masked area, thus improving the quality of generated images for anomaly detection. However, the use of inpainting for anomaly detection has not been extensively explored in BML detection applications.

In this paper, we propose a semi-supervised anomaly detection framework based on mask inpainting techniques to detect BMLs, as illustrated in Figure 1. Unlike a conventional UAD system, this approach requires a femur bone mask as ROI for training on healthy images with masked regions. A generative model is then trained to inpaint the masked region. During prediction, an input image with anomalies in the femur bone is translated to one with a healthy femur bone region. Post-processing steps are applied to generate the final BML segmentation. The proposed pipeline is not fully unsupervised due to the need for femur bone masks. However, femur bone segmentation is an easy task due to its simple and regular shapes as demonstrated by high Dice coefficient (96.94%) in our previous work[8]. Our proposed method avoids detecting anomalies on the full image and instead focuses on specific regions, preventing information loss that occurs when images are resampled to a lower resolution due to the difficulties in generating high-resolution images.

## 2. METHOD

In this work, we used an open dataset from the Osteoarthritis Initiative (OAI) sponsored by the National Institutes of Health. The knee MRI dataset comprises images from 210 subjects for training, each containing approximately 36 knee intermediate weighted fat-suppressed (IWFS) image slices, as well as images from 45 subjects for validation and 45 subjects for testing. The femur bone masks for the training set and the BML masks for the validation and test sets were annotated by trained medical professionals using a semi-automated tool.

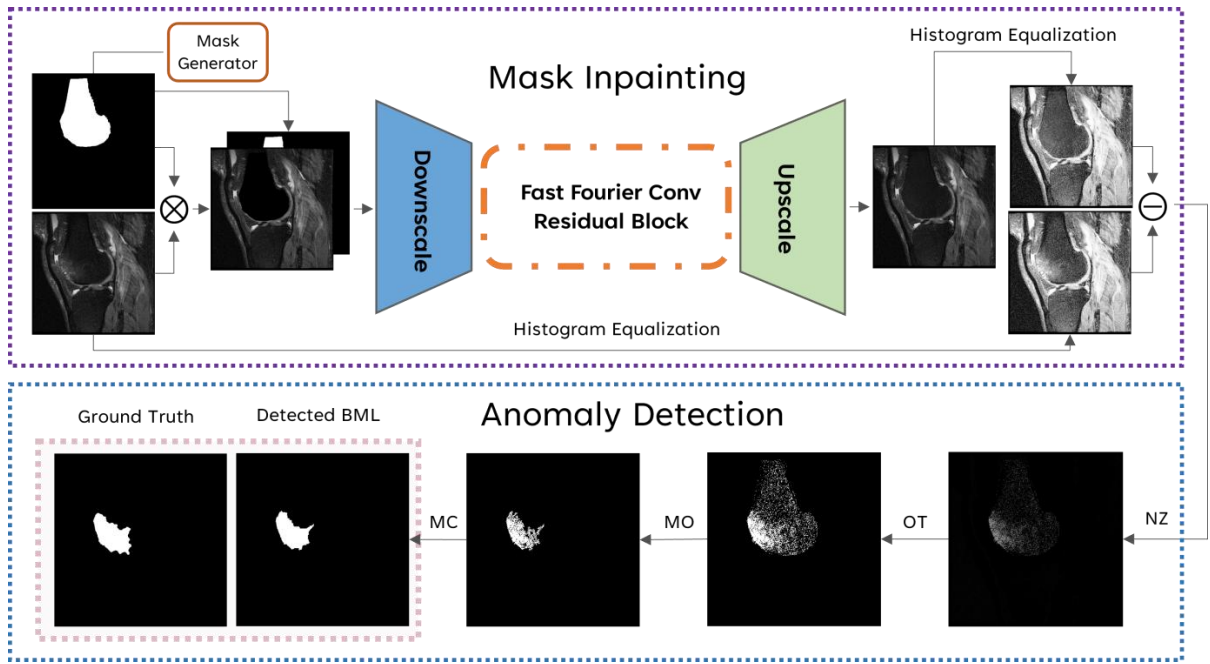

Figure 1. The proposed BML detection pipeline with two main steps: mask inpainting and anomaly detection. NZ: negative zeroed out; OT: Otsu thresholding; MO: morphological opening; MC: morphological closing.

To generate the femur bone mask, we applied U-Net++[9] as a one-stage 3D semantic segmentation model. This model was trained on a dedicated training set, utilizing data augmentation techniques to enhance performance. Specifically, the 3D volumes were randomly cropped into 256x256x16 patches. During testing, the masks were generated in 3D and subsequently split into 2D slices for further processing. This step achieved an ideal segmentation performance with an average Dice coefficient of 94.34%. For inpainting, we utilized a resolution-robust large mask inpainting model (LaMa)[10] based on a GAN structure with a generator and a discriminator. The generator employs fast Fourier convolutions (FFCs)[11] to expand the receptive field. FFC is based on channel-wise fast Fourier transform (FFT) which allows the model to have a wide receptive field from the early layers and enables efficiency of high-resolution images inpainting. During training, the 3D images were split into 2D slices, and we selected slices without BMLs paired with their manually annotated femur bone masks. Thus, the LaMa model was trained on 2D slices from normal knee MRIs without BMLs. Each slice was resized to different resolutions for various experiments. The input image includes a manually annotated bone mask and a knee MRI with the femur bone masked, resulting in a two-channel input. Data augmentation techniques, such as random vertical and horizontal flipping, and random bias field[12] augmentation, were applied to enhance model robustness.

During testing, the input image $x$ is masked by the generated bone mask $m$ and get masked image $x \otimes m$. Stacked with mask $m$, we have two-channel input: $stack(x \otimes m, m)$. We applied histogram equalization to enhance global contrast of input $x$ and output $\tilde{x}$, and get enhanced images $x' = HE(x)$ and $\tilde{x}' = HE(\tilde{x})$. Next, we performed a subtraction between the reconstructed and original images to create a difference map. As BMLs tend to appear brighter than the surrounding marrow, we zeroed out negative values, thus getting the final difference map $D(x', \tilde{x}') = |\tilde{x}' - x'|$. To generate binary anomaly mask, each output slice was processed using Otsu's method[13]. After Otsu's thresholding, we get an initial binary mask $m$. Due to the presence of noise in knee MRIs, the subtraction operation cannot effectively locate these noises, resulting in the mask $m$ containing some very small and discontinuous structures. Therefore, a morphological opening filter is used to remove small structural noise from the binary mask and a morphological closing is applied to fill small holes in the detected BMLs, refining the final anomaly mask. Notably, all the hyperparameters were determined using a validation dataset.

## 3. RESULTS

Table 1. Quantitative comparison of the detected masks on different input image resolutions using the proposed method and the comparison MKD method.

| Method | Image Size | Dice | IoU | Sensitivity | Specificity | Accuracy |
|---|---|---|---|---|---|---|
| Ours | 128 × 128 | 0.32 | 0.21 | **0.65** | 0.87 | 0.85 |
|  | 192 × 192 | 0.35 | 0.24 | 0.63 | 0.91 | 0.88 |
|  | 256 × 256 | 0.39 | 0.27 | 0.59 | 0.92 | 0.90 |
|  | 320 × 320 | 0.38 | 0.27 | 0.59 | 0.91 | 0.90 |
|  | 448 × 448 | **0.42** | **0.30** | 0.60 | **0.94** | **0.91** |
| MKD | 320 × 320 | 0.15 | 0.10 | 0.79 | 0.48 | 0.49 |
|  | 448 × 448 | 0.20 | 0.12 | 0.68 | 0.70 | 0.69 |

To test the proposed method's ability to predict high-resolution images when trained on low-resolution data, we resampled the original MRI images to different resolutions. We compared the results using average Dice score, average Intersection over Union (IoU), pixel-level sensitivity, specificity, and accuracy. Additionally, we compared our method with the state-of-the-art anomaly detection method MKD, which has achieved high performance in various medical anomaly detection tasks[14]. Quantitative comparison results between our method and MKD are shown in Table 1. The overall results in Dice and IoU are low, not only because BML segmentation is challenging due to their small size and low contrast from the surrounding marrow, but also because we did not remove other anomalies besides BML. It also indicates that global detection methods such as MKD struggle on detection on high-resolution images. Even when using bone masks to constrain the location of detected anomalies, MKD results were unsatisfactory. Our method outperforms MKD, indicating that higher-resolution images improve segmentation performance, as reflected by higher Dice and IoU scores. Specifically, the original image size of 448x448 achieved the highest Dice of 0.42 and IoU of 0.30, compared to the lowest input size of 128x128 that only achieved a Dice of 0.32 and IoU of 0.21. Sensitivity showed limitations with around 60%, as BML is a low-contrast lesion in IWFS MRI. This issue is particularly prominent near the boundary between BML and surrounding marrow, leading to a poor overlap between the predicted BML mask and the manually labeled mask, especially for small BMLs with a large portion on the boundary. Although the quantitative results are limited, the qualitative results shown in Figure 2 demonstrate that our method can accurately localize BMLs, even when they are small, as seen in case 1. We separated the test data into five groups based on BML area, which indicate that different BML sizes yield varying outcomes. As shown in Figure 3, with increasing BML size, both Dice and IoU scores improved, as the proportion of distinguishable boundary decreases.

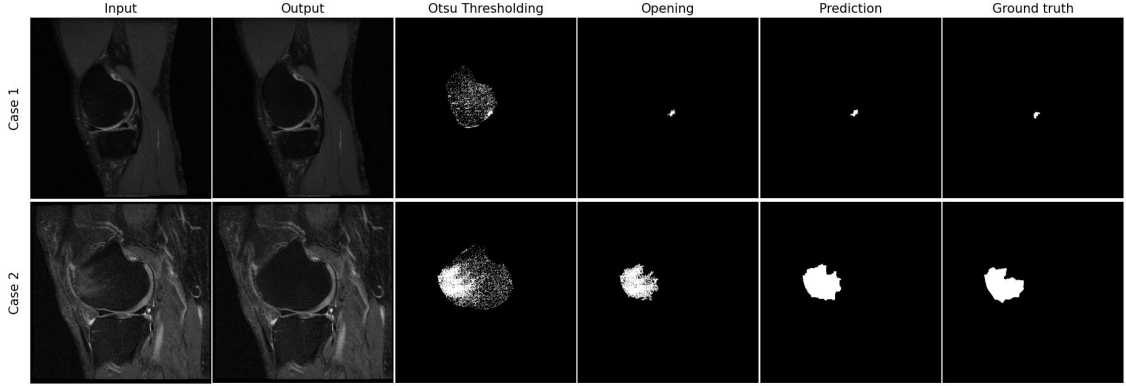

Figure 2. Sample cases showing pixel-level localization of anomalous image regions. The third and fourth columns show the results after Otsu thresholding and morphological opening operations in the proposed pipeline. The fifth column shows the results after applying morphological closing.

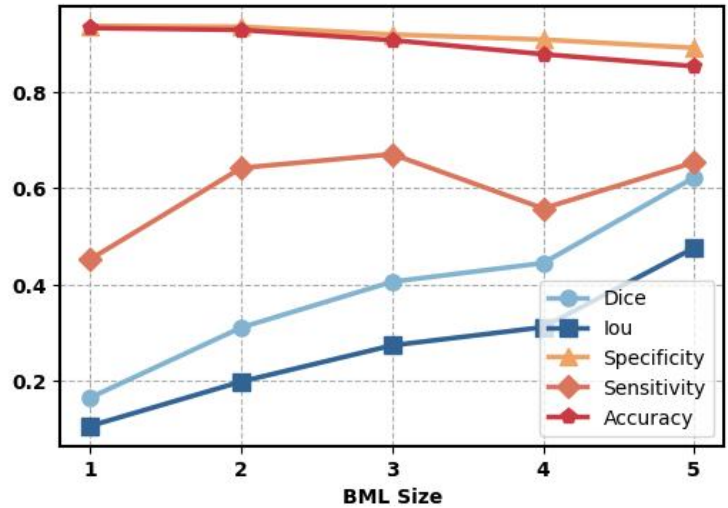

Figure 3. Performance of our proposed method on different BML. The X-axis represents the relative size of the BMLs.

## 4. CONCLUSIONS

In this paper, we presented a semi-supervised anomaly detection system for detecting BMLs in high-resolution knee MRIs. Utilizing a combination of a high-performance 3D femur bone segmentation model, a mask inpainting model, and post-processing techniques, we aimed to address the challenges of segmenting BMLs, which are often small and difficult to distinguish from the surrounding bone marrow. The method was evaluated using knee MRIs at various resolutions, showing an overall advantage over the MKD, especially at segmenting higher-resolution images. The proposed method has the potential to improve BML detection, and the detected masks could serve as valuable markers for future tasks such as 3D segmentation or classification.

## 5. NEW WORK TO BE PRESENTED

In the full manuscript, we will provide a more comprehensive discussion of our results, including a detailed comparison with other anomaly detection methods to highlight why our approach is superior. We will provide more examples, both visually and quantitatively, on the results using our method and the comparison method. We will also analyze the performance differences in depth and support our findings with extensive visualizations.

## ACKNOWLEDGEMENTS

This work was supported by the National Research Foundation (NRF) of Korea funded by the Ministry of Science and ICT (RS-2024-00338438). This work is not being, and has not been, submitted for publication or presentation elsewhere.